# Distinct neurodynamics of functional brain networks in Alzheimer's disease and frontotemporal dementia as revealed by EEG

Sungwoo Ahn[1,2], Evie A. Malaia[3], Leonid L Rubchinsky[4,5,*]

[1]Department of Mathematics, East Carolina University, Greenville, NC
[2]Alliance for Brain Stimulation, College of Allied Health Sciences, East Carolina University, Greenville, NC
[3]Department of Speech, Language and Hearing, University of Alabama, Tuscaloosa, AL
[4]Department of Mathematical Sciences, Indiana University, Indianapolis, IN
[5]Stark Neurosciences Research Institute, Indiana University School of Medicine, Indianapolis, IN

*Corresponding author:

Leonid Rubchinsky

Department of Mathematical Sciences
Indiana University Indianapolis
402 N. Blackford St

Indianapolis, IN 46202

lrubchin@iu.edu

**Abstract**

*Objective*
While Alzheimer's disease (AD) and frontotemporal dementia (FTD) show some common memory deficits, these two disorders show partially overlapping complex spatiotemporal patterns of neural dynamics. The objective of this study is to characterize these patterns to better understand the general principles of neurodynamics in these conditions.

*Methods*
A comprehensive array of methods to study brain rhythms and functional brain networks are used in the study, from spectral power measures to Lyapunov exponent, phase synchronization, temporal synchrony patterns, and measures of the functional brain connectivity. Furthermore, machine learning techniques for classification are used to augment the methodology.

*Results*
Multiple measures (spectral, synchrony, functional network organization) indicate an array of differences between neurodynamics between AD and FTD, and control subjects across different frequency bands.

*Conclusions*
These differences taken together in an integrative way suggest that AD neural activity may be less coordinated and less connected across areas, and more random, while FTD shows more coordinated neural activity (except slow frontal activity).

*Significance*
AD and FTD may represent opposite changes from normal brain function in terms of the spatiotemporal coordination of neural activity. Deviations from normal in both directions may lead to neurological deficits, which are specific to each of the disorders.

**Highlights**

- Neurodynamics in AD and FTD are complex and require multiple neural activity measures.
- AD shows less coordinated, more random activity, while FTD exhibits greater neural coordination than controls.
- AD and FTD represent distinct neurodegeneration patterns in spatiotemporal coordination of neural activity.

# 1. Introduction

Distinguishing between Alzheimer's disease (AD) and frontotemporal dementia (FTD) through clinical symptoms alone can be challenging, particularly in early stages of either disorder. Since both conditions involve progressive neurodegeneration, but differ in the mechanisms of it, understanding disease-specific changes to neural dynamics using data with high temporal resolution – such as electroencephalography (EEG) - could both provide insights into disease mechanisms, and, potentially, help develop multimodal (neuro-behavioral) diagnostic markers for differential diagnostics. In this work, we analyzed the dynamics in EEG patterns in patients with Alzheimer's disease (AD) and frontotemporal dementia (FTD) as compared to control participants, aiming to understand the functional networks of the brain and their link to behavior in clinical populations.

Alzheimer's disease (AD) is the most prevalent neurodegenerative disorder, affecting approximately 10.9% of individuals aged 65 and older (Alzheimer's Association, 2024). The primary symptom of AD is progressive memory loss, which begins with episodic memory deficits affecting recent events, advances to impairment in word retrieval, then procedural memory (i.e. motor and executive functions), and finally results in retrograde amnesia, disrupting long-term autobiographical memories. In terms of pathophysiology, AD has been linked to the extracellular buildup of insoluble amyloid-β (Aβ) plaques and the intracellular aggregation of hyperphosphorylated tau proteins, forming neurofibrillary tangles (Bloom, 2014). These pathological changes lead to widespread neuronal death, synaptic loss, and brain atrophy, with structural damage progressing unevenly across different brain regions (Guzmán-Vélez et al., 2022). The precise neurofunctional mechanisms connecting AD pathophysiology to cognitive decline remain unclear, especially as the extent of neurodegeneration does not consistently align with the severity of cognitive impairment (Arbabyazd et al., 2023).

Frontotemporal dementia (FTD) is the second most common cause of early-onset dementia, which affects younger demographics: it typically manifests between ages 45 and 65. FTD is not a single disorder, like AD; rather, it is a group of neurodegenerative disorders characterized by progressive atrophy of the frontal and temporal lobes. Unlike Alzheimer's disease, where memory impairment is primary, FTD initially disrupts social cognition and executive function. Clinically, advanced FTD presents as profound changes in personality, behavior, and language, including disinhibition, apathy, compulsive behaviors, and difficulties in both speech production and comprehension. The underlying pathophysiology involves abnormal accumulations of tau and TDP-43 proteins, which leads to widespread neuronal loss (Mann & Snowden, 2017). Recent neuroimaging studies using functional connectivity analyses have demonstrated alterations in large-scale brain networks that correspond to the clinical symptoms of FTD, with dynamic connectivity changes influencing the progression of cognitive and behavioral impairments (Agosta et al., 2023; Ferreira et al., 2022; Salcini et al., 2020). These findings highlight the need to investigate disruptions in neural network dynamics, rather than structural degeneration alone.

Functional connectivity neuroimaging analyses focused on dynamics of brain activity have already helped reveal distinct patterns of changes in spatiotemporal dynamics of AD and FTD. For example, in AD, fMRI data analysis showed progressive loss of high-order neural interactions, which led to disorganized and "bursty" functional connectivity, especially in cingulate and limbic networks (Arbabyazd et al., 2023). FTD, on the other hand, has been found to exhibit more localized disruptions in frontotemporal network coordination (Chu et al., 2024). EEG data, with its high temporal resolution, analyzed using measures of non-linear dynamics, can offer further insight into differences in between AD and FTD. While both disorders exhibit progressive neurodegeneration, their cognitive and behavioral manifestations arise from distinct disruptions in functional dynamics of neural activity. Use of non-linear metrics can help understand how global network reconfigurations differ between these two conditions, potentially supporting diagnostic differentiation and advancing our understanding of distinctive neural mechanisms of AD and FTD.

To approach this issue, we are using several metrics, that are focused on the dynamics and structure of the functional network as recorded by EEG signals. We use standard spectral measures that characterize power spectra in different spectral bands (theta, alpha, beta, and low gamma). We use Lyapunov exponent, that characterizes the stability/instability of dynamical system trajectories as they move along possible dynamical states. We use measures of phase synchronization and of its temporal patterning that characterize the properties of temporal coordination of activity between network nodes. We also use clustering and global efficiency of synchrony-based functional brain networks – measures that describe the results of temporally sensitive parameter changes on spatial organization of neural function. To complement conventional statistical analyses assessing salience of proposed measures to characterizing neural dynamics in the two clinical cohorts, we also apply machine learning analyses, which are not bound by the same assumptions as parametric statistics, but it can be sensitive to the structure of the underlying manifold, even if it is characterized by a limited set of parameters. The combination of these approaches provides a powerful framework for exploring complex, multi-scale spatiotemporal patterns of neural activity. By adopting a dynamical systems perspective, we seek to provide insights into the functional connectivity patterns and neural synchronization dynamics that characterize AD and FTD.

Prior research indicates that the differences of neural dynamics among AD, FTD, and control that are discussed above are relatively small in quantitative terms (Wang et al., 2024; Zheng et al., 2023). Thus, we are trying to incorporate multiple measures of network dynamics, so that we can consider the outcomes of the analysis of individual measures in an integrative manner to make advancements in comprehensive understanding of the neural mechanistic underpinning of the disease. Taking multiple, potentially small, dynamics differences together let us come to more confident conclusions about the underlying functional networks. Furthermore, contrasting neural dynamics in AD and FTD not only with controls, but also with each other, provides a more heterogeneous set of different neuronal networks for analysis and may help to elucidate potentially subtle neurophysiological differences. While there may be some diagnostic applications, our primary objective is to explore neurophysiological mechanisms. The multiple measures of neural dynamics and organization of the functional EEG-based networks taken together can help to understand what kind of spatiotemporal coordination of neural activity on different temporal and spatial scales may stand behind the studied disorders.

## 2. Methods

## 2.1. Subjects and experimental data

We used 19-channel electroencephalogram (EEG) data based on the international 10-20 system, recorded from 36 Alzheimer's disease (AD) subjects, 23 frontotemporal dementia (FTD) subjects, and 29 healthy control (CT) subjects. This dataset is publicly available on OpenNeuro (https://openneuro.org/datasets/ds004504/versions/1.0.7) (Miltiadous et al., 2023). A summary of the participant data is provided in Table 1, while a detailed description of the dataset can be found in the referenced paper. Here, mental state was evaluated by the international Mini-Mental State Examination (MMSE). The detailed values for each subject is available in (Miltiadous et al., 2023).

**Table 1.** Dataset description

| Group | # of subjects | Gender (M/F) | Age (mean±sd) | MMSE (mean±sd) | EEG duration in minutes Mean (min, max) |
|-------|---------------|--------------|---------------|----------------|------------------------------------------|
| AD | 36 | 12/24 | 66.4±7.9 | 17.75±4.5 | 13.5 (min=5.1, max=21.3) |
| FTD | 23 | 14/9 | 63.6±8.2 | 22.17±8.22 | 12 (min=7.9, max=16.9) |
| CT | 29 | 18/11 | 67.9±5.4 | 30±0 | 13.8 (min=12.5, max=16.5) |

The eyes-closed resting state EEG signals were recorded using a clinical EEG device (Nihon Kohden 2100) with 19 scalp electrodes (Fp1, Fp2, F7, F3, Fz, F4, F8, T3, C3, Cz, C4, T4, T5, P3, Pz, P4, T6, O1, O2) and two reference electrodes (A1 and A2) placed on the mastoids for impedance checks. The sampling rate was 500 Hz, and the resolution was 10 µV/mm.

## 2.2 Data processing

We used the preprocessed dataset (referred to as 'derivatives'), which was initially filtered and pre-processed using EEGLAB by (Miltiadous et al., 2023). A brief description of their process is provided below, while the detailed procedure can be found in the referenced paper.

To remove noise and potential artifacts such as eye and jaw artifacts, the EEG signals were re-referenced to the average value of A1 and A2. A Butterworth band-pass filter (0.5-45 Hz) was then applied. The signals were further processed using artifact subspace reconstruction (ASR), followed by independent component analysis (ICA) with the RunICA algorithm. The automatic classification method 'ICLabel' within EEGLAB routines was used for artifact classification.

To systematically analyze the data (due to the different time-durations of EEG recordings) in the following analysis, the continuous EEG signals were divided into several nonoverlapping 60 sec time windows for further analysis with custom Matlab software (MATLAB R2024b). These 60 sec time windows give enough oscillatory cycles to compute the time-series measures even at low-frequency bands. All time-series measures below in Sec 2.3 were computed for each individual electrode or each electrode pair in 60 sec nonoverlapping time windows, and then each measure was averaged over the whole time period per electrode or per pair.

## 2.3. Data analysis

### 2.3.1. Spectral measures
For the purposes of data analysis, the spectral bands were defined as follows: theta (4-7 Hz), alpha (8-12 Hz), beta (13-30 Hz), and low gamma (31-45 Hz). Power spectral density (PSD) from each electrode was computed using the built-in Matlab *pwelch* function for Welch's power spectral density estimate using 50% overlapped Hamming windows. This PSD is referred to here as the spectral power.

In addition to the spectral power analysis described above, we also utilized irregular-resampling auto-spectral analysis (IRASA; Wen & Liu, 2016) to separate fractal (1/f) and oscillatory components. The resulting spectral power characteristics of the oscillatory component are referred to here as the 1/f oscillatory power.

### 2.3.2. Measure of dynamics regularity

To characterize the regularity of neural dynamics we used the maximal Lyapunov exponent, which characterizes the rate of exponential divergence of close trajectories in the reconstructed phase space (Abarbanel 1996; Kantz & Schreiber 2003). A larger Lyapunov exponent indicates a larger divergence rate, which accompanies the chaoticity of dynamics. We estimated the largest Lyapunov exponent using the built-in Matlab *lyapunovExponent* function with time lag=10 ms, dim=5, and the expansion range=100 ms. Since the Lyapunov exponent would rather be computed over the full (broadband) signal, and the data were band-stopped at 45Hz, the results with Lyapunov exponent should probably be considered as not very confident.

### 2.3.3. Synchronized dynamics measures

The synchronization analysis methods used here were described previously in detail (Ahn et al., 2014; Ahn & Rubchinsky, 2013; Park et al., 2010). Briefly, signals were Kaiser windowed and digitally filtered using a finite impulse response filter in four frequency (theta, alpha, beta, and low gamma) bands. The Hilbert transform was used to reconstruct phases of oscillations. The reconstructed phases were used to estimate the synchronization (phase-locking) index:

$$\gamma = \left\| \frac{1}{N} \sum_{j=1}^{N} e^{i\theta(t_j)} \right\|^2,$$

where $\theta(t_j)$ is the difference of the phases of oscillations at the given time point $t_j$. $N$ is the total number of such time points. This phase-locking index $\gamma$ varies from 0 to 1 (perfect phase synchronization) and detects phase locking with any (not necessarily zero) phase lag (Hurtado et al., 2004; Pikovsky et al., 2001).

The phase-locking index $\gamma$ characterizes the synchronization strength averaged over the analysis window. If the oscillations are synchronized on average, then it is possible to check whether oscillations are in the synchronized state at a specific cycle of oscillations (Ahn et al., 2011, 2014; Ahn & Rubchinsky, 2013; Park et al., 2010). We extract intervals during which the phase difference is close to the preferred value and the intervals during which the phase difference substantially deviates from the preferred value (desynchronizations). Whenever the phase of one signal crosses zero level from negative to positive values, we record the phase of the other signal, generating a set of consecutive values $\{\phi_i\}$, $i=1,..., M$. These $\phi_i$ represent the phase difference between two signals when the phase of one signal crosses zero. After determining the most frequent value of $\phi_i$, all the phases are shifted accordingly (for different episodes under consideration) so that averaging across different episodes (with potentially different phase shifts) is possible. Thus, this approach is not concerned with the value of the phase shift between signals, but rather with the maintenance of the constant phase shift (phase-locking).

Temporal dynamics are considered to be desynchronized if the phase difference deviates from the preferred phase difference by more than $\pi/2$ as in the earlier studies. The duration of the desynchronized episodes is measured in cycles of the oscillations, which allows for the comparison of temporal patterns of synchronization between different brain rhythms. This approach considers the maintenance of the phase difference in time and distinguishes between many short desynchronizations, few long desynchronizations, and possibilities in between even if they have the same average synchrony strength. This is quantified with the desynchronization ratio (DR): the ratio of the relative frequencies of the desynchronizations lasting for one cycle to longer than 4 cycles of oscillations (Ahn et al., 2014; Malaia et al., 2020). Smaller DR points to longer desynchronizations while larger DR points to shorter desynchronizations. Patterning of desynchronizations can vary independently of the average synchronization strength γ (Ahn et al., 2014; Ahn & Rubchinsky, 2017; Nguyen & Rubchinsky, 2021, 2024) and may be more sensitive to the changes in the synchronous dynamics than changes in the synchronization strength (Ahn et al., 2017).

### 2.3.4. Functional network measures

We applied graph theory measures to networks defined as systems consisting of a set of nodes (representing electrodes) linked by edges (representing functional interactions). Initially, we constructed a symmetric connectivity matrix $A=[a_{ij}]$ (19 by 19 matrix) where the diagonal elements were set to zero. Here, each element $a_{ij}$ represents the phase-locking index γ between i-th and j-th nodes. To convert the matrix A into a binary matrix B, we used a thresholding approach. In the matrix B, a value of 0 represented no connection, and 1 represented a connection. We determined the threshold by using values of the phase-locking index γ from all control subjects. Specifically, we tested a threshold value as the mean value of the phase-locking index from all control subjects at the given frequency band.

Two kinds of graph theory measures were calculated based on the binary matrix B. The clustering coefficient (CC) quantifies the tendency of nodes in a graph to form clusters or tightly interconnected groups. A higher clustering coefficient indicates a higher density of triangles or closed connections among connected nodes, reflecting a more segregated network structure (Bullmore & Sporns, 2012; Latora & Marchiori, 2001; Watts & Strogatz, 1998). The clustering coefficient of a node $v$ is calculated as the fraction of closed triangles among three nodes. Specifically, if $deg(v)$ is the degree of a given node $v$ and $T(v)$ is the number of closed triangles centered at the node $v$ then, $CC(v)$ is defined as

$$CC(v) = \frac{T(v)}{\deg(v) \times (\deg(v)-1)/2}.$$

This measure captures the density of connections and the tendency of the connectedness of the node $v$ to be interconnected with other nodes.

Furthermore, the characteristic of network efficiency is considered. The efficiency for each node assesses the connectedness of spatially distant regions. The efficiency for each node was computed as the mean value of the reciprocal shortest path lengths between the given node and all other nodes in the network. Specifically, if $d(v, u)$ represents the shortest path length (minimum number of edges in any path) between node $v$ and node $u$, then $EF(v)$, the efficiency of node $v$, is defined as the mean value of all $1/d(v, u)$ values for $v \neq u \in V$, where $V$ is the set of all nodes in the network. If there is no path from $v$ to $u$, then $1/d(v, u)$ is defined as zero. High efficiency may promote processing and effective information integration across the network.

These two measures may provide valuable insights into the brain's interconnectedness and the efficiency of information segregation and propagation through the network (Bullmore & Sporns, 2012; Latora & Marchiori, 2001; McDonnell et al., 2021; Micheloyannis et al., 2006; Stam & Reijneveld, 2007; Watts & Strogatz, 1998).

### 2.3.5. Statistical data analysis

For each electrode (or each pair) for each subject, we first computed the above neural dynamical measures (PSD, 1/f oscillatory power, Lyapunov exponent, phase-locking index, DR, CC, and EF) over every 60 sec non-overlapping windows (note that there were several of these 60 sec non-overlapping windows). Subsequently, each of these measures was averaged across all non-overlapping windows.

To explore the effects of the locations, we created two regions of interest (ROIs) including channels in the frontal region (Fp1, Fp2, F7, F3, Fz, F4, F8) and the posterior region (P3, Pz, P4, T5, T6, O1, O2). By using two ROIs, one can explore the effect of the different brain regions, but one can reduce the dimensionality of the analysis while enhancing statistical power.

All comparisons were first subjected to a mixed design repeated measures of ANOVA (between-subject factor as a Group; within-subject factor as a ROI) with significance level at $p<0.05$. If there is any significant effect of ROI or Group*ROI interaction, then we further perform the post-hoc test (Tukey's HSD) with one-way ANOVA to explore the group difference at each ROI with significance level at $p<0.05$. Before using the mixed design repeated measures of ANOVA, sphericity was tested by using Mauchly's test. When the assumption of sphericity was violated, the degree of freedom in Greenhouse-Geisser correction was used.

### 2.3.6 Classification

We used a subset of the above time series measures (1/f oscillatory power, PSD, Lyapunov exponent, phase-locking (synchronization) index, DR, CC, and EF) of each electrode or each pair of electrodes to explore whether these measures can classify different conditions based on the values of measures. We consider two different classifications processes. First, we compare AD+FTD with healthy control subjects to identify illness. Second, we compare AD with FTD to identify different disease status. To select a subset to use in the machine learning classification algorithms, we first performed the one-way ANOVA (between-subject factor as a Group) for each measure to extract the significant features. To reduce the number of possible combinations in the machine learning classification algorithms, we used the relatively small subset by letting the significant p-value to be less than 0.0005 from the one-way ANOVA. We further restricted the number of features as at most 3-6 features into the classification algorithms to reduce the computation time and resource. Due to the smaller sample size of the group, data values of all participants were used for training the classifiers. To prevent the overfitting, we used the tenfold cross-validation method.

The following five machine learning algorithms were employed: medium tree, linear discriminant, quadratic support vector machine, fine K-nearest neighbors, and bagged tree. The accuracy, sensitivity, and specificity of the classification algorithms were calculated by the following formulae:

Accuracy $= \frac{TP+TN}{TP+TN+FP+FN}$

Sensitivity $= \frac{TP}{TP+FN}$

Specificity $= \frac{TN}{TN+FP}$

|  | | True Class | |
|---|---|---|---|
|  | | Positive | Negative |
| Predicted | Positive | TP | FP |
| Class | Negative | FN | TN |

where TP is true positive, FP is false positive, FN is false negative, and TN is true negative.

# 3. Results

## 3.1. Behavioral measures

We performed one-way ANOVA (between-subjects factor as a Group) to explore the group differences for MMSE scores (see Table 1 for the summary). There was a significant main effect of Group ($F_{(2, 85)}$=119.7, p<1.0e-16). The post-hoc test showed that MMSE of CT subjects was significantly higher than those of AD subjects (Tukey's HSD, p=5.10e-9) and FTD subjects (Tukey's HSD, p=5.10e-9). MMSE of FTD subjects was significantly higher than that of AD subjects (Tukey's HSD, p=3.97e-6).

## 3.2. Spectral measures

### 3.2.1. 1/f Oscillatory Power

At theta band oscillatory power, there were significant main effects of Group and ROI as well as a significant effect of Group*ROI interaction (see Table 2 for the detail statistical values). The post-hoc test in the frontal region showed that 1/f oscillatory power of AD subjects was significantly higher than that of CT subjects (p=1.64e-3, Fig. 1). The post-hoc test in the posterior region showed that 1/f oscillatory power of CT subjects was significantly lower than those of AD subjects (p=3.75e-5, Fig. 1) and FTD subjects (p=3.83e-3, Fig. 1).

At alpha band oscillatory power, there were significant main effects of Group and ROI as well as a significant effect of Group*ROI interaction (Table 2). The post-hoc test in the frontal region showed that 1/f oscillatory power of CT subjects was significantly higher than those of AD subjects (p=1.17e-2, Fig. 1) and FTD subjects (p=4.80e-3, Fig. 1). The post-hoc test in the posterior region showed that 1/f oscillatory power of CT subjects was significantly higher than those of AD subjects (p=2.66e-5, Fig. 1) and FTD subjects (p=1.21e-3, Fig. 1).

At beta band oscillatory power, there were no main effects of Group and ROI as well as no effect of Group*ROI interaction (Table 2).

At low gamma band oscillatory power, there were a significant main effect of ROI and a significant effect of Group*ROI interaction, but no effect of Group (Table 2). The post-hoc test in the frontal region showed that 1/f oscillatory power of FTD subjects was significantly higher than that of CT subjects (p=2.98e-2, Fig. 1). The post-hoc test in the posterior region showed no differences among groups (p>0.05).

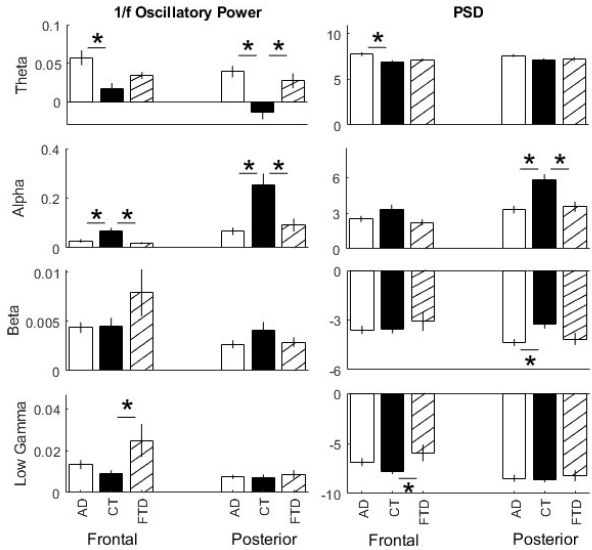

**Figure 1.** 1/f Oscillatory power and PSD. AD represents Alzheimer's Disease. CT represents Control. FTD represents Frontotemporal Dementia. Mean±SEM. Tukey's HSD post-hoc test. * p<0.05.

**Table 2.** 1/f Oscillatory power

|  | Theta band | Alpha band | Beta band | Low Gamma band |
|---|---|---|---|---|
| Group | F(2,85)=9.3 p=2.23e-4 | F(2,85)=11.5 p=3.69e-5 | F(2,85)=0.3 p>0.05 | F(2,85)=2.5 p>0.05 |
| ROI | F(1,85)=41.8 p=6.05e-9 | F(1,85)=55.4 p=7.38e-11 | F(1,85)=0.0 p>0.05 | F(1,85)=21.6 p=1.20e-5 |
| Group*ROI | F(2,85)=5.3 p=6.98e-3 | F(2,85)=12.4 p=1.95e-5 | F(2,85)=1.0 p>0.05 | F(2,85)=5.4 p=6.47e-3 |

*3.2.2. PSD*

At theta band PSD, there were a significant main effect of Group and a significant effect of Group*ROI interaction, but no effect of ROI (Table 3). The post-hoc test in the frontal region showed that PSD of AD subjects was significantly higher than that of CT subjects (p=4.68e-3, Fig. 1). The post-hoc test in the posterior region showed no differences among groups (p>0.05).

At alpha band PSD, there were significant main effects of Group and ROI as well as a significant effect of Group*ROI interaction (Table 3). The post-hoc test in the frontal region showed no differences among groups (p>0.05). The post-hoc test in the posterior region showed that PSD of CT subjects was significantly higher than those of AD subjects (p=2.70e-5, Fig. 1) and FTD subjects (p=7.40e-4, Fig. 1).

At beta band PSD, there were a significant main effect of ROI and a significant effect of Group*ROI interaction, but no effect of Group (Table 3). The post-hoc test in the frontal region showed no differences among groups (p>0.05). The post-hoc test in the frontal region showed no differences among groups (p>0.05). The post-hoc test in the posterior region showed that PSD of CT subjects was significantly higher than that of AD subjects (p=7.73e-3, Fig. 1).

At low gamma PSD, there were a significant main effect of ROI and a significant effect of Group*ROI interaction, but no effect of Group (Table 3). The post-hoc test in the frontal region showed that PSD of FTD was significantly higher than that of CT subjects (p=4.32e-2, Fig. 1). The post-hoc test in the posterior region showed no difference among groups (p>0.05).

**Table 3.** PSD

|          | Theta band              | Alpha band              | Beta band               | Low Gamma band          |
|----------|-------------------------|-------------------------|-------------------------|-------------------------|
| Group    | F(2,85)=4.1<br>p=1.93e-2 | F(2,85)=8.1<br>p=5.93e-4 | F(2,85)=1.1<br>p>0.05   | F(2,85)=1.5<br>p>0.05   |
| ROI      | F(1,85)=1.1<br>p>0.05   | F(1,85)=142.9<br>p<1.0e-16 | F(1,85)=20.2<br>p=2.18e-5 | F(1,85)=95.2<br>p=1.60e-15 |
| Group*ROI | F(2,85)=4.6<br>p=1.23e-2 | F(2,85)=17.4<br>p=4.79e-7 | F(2,85)=12.7<br>p=1.44e-5 | F(2,85)=6.2<br>p=3.00e-3 |

### 3.3. Dynamics regularity measure

*Lyapunov exponent*
There were a significant main effect of ROI and a significant effect of Group*ROI interaction, but no effect of Group (Table 4). The post-hoc test in the frontal region showed no difference among groups (p>0.05). The post-hoc test in the posterior region showed that Lyapunov exponent of AD subjects was significantly higher than that of CT subjects (p=8.04e-3, Fig. 2).

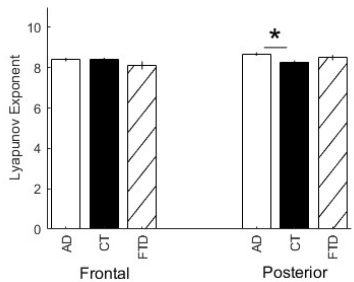

**Figure 2.** Lyapunov exponent. Mean±SEM. Tukey's HSD post-hoc test. *:p<0.05.

**Table 4.** Lyapunov exponent

| Group     | F(2,85)=1.6<br>p>0.05   |
|-----------|-------------------------|
| ROI       | F(1,85)=14.0<br>p=3.36e-4 |
| Group*ROI | F(2,85)=11.1<br>p=5.28e-5 |

### 3.4. Synchronized dynamics measures

*3.4.1. Phase synchronization index*
At theta band, there were a significant main effect of Group and a significant effect of Group*ROI interaction, but no effect of ROI (Table 5). The post-hoc test in the frontal region showed that the phase synchronization index of CT subjects was significantly higher than that of AD subjects (p=1.02e-4, Fig. 3). The post-hoc test in the posterior region showed that the phase synchronization index of CT subjects was significantly higher than that of AD subjects (p=1.81e-3, Fig. 3).

At alpha band, there were significant main effects of Group and ROI as well as a significant effect of Group*ROI interaction (Table 5). The post-hoc test in the frontal region showed no differences among groups (p>0.05). The post-hoc test in the posterior region showed that the phase synchronization index of CT subjects was significantly lower than that of FTD subjects (p=7.78e-3, Fig. 3).

At beta band, there were significant main effects of Group and ROI as well as a significant effect of Group*ROI interaction (Table 5). The post-hoc test in the frontal region showed no differences among groups (p>0.05). The post-hoc test in the posterior region showed that the phase synchronization index of FTD subjects was significantly higher than CT subjects (p=1.13e-3, Fig. 3).

At low gamma band, there were significant main effects of Group and ROI as well as a significant effect of Group*ROI interaction (Table 5). The post-hoc test in the frontal region showed that the phase synchronization index of AD subjects was significantly lower than that of FTD subjects (p=1.58e-2, Fig. 3). The post-hoc test in the posterior region showed that the phase synchronization index of AD subjects was significantly lower than that of FTD subjects (p=1.64e-2, Fig. 3).

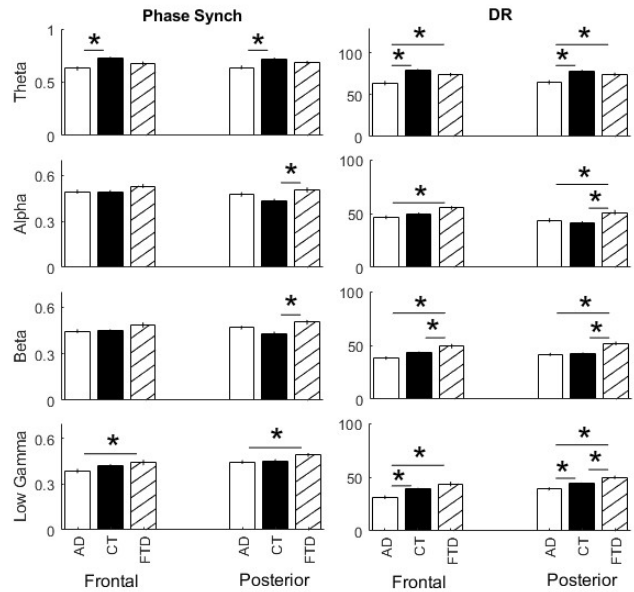

**Figure 3.** Phase synchronization index and desynchronization ratio (DR). Mean±SEM. Tukey's HSD post-hoc test. *: p<0.05.

**Table 5.** Phase synchronization index

|  | Theta band | Alpha band | Beta band | Low Gamma band |
|---|---|---|---|---|
| Group | $F_{(2,85)}=8.0$ p=6.38e-4 | $F_{(2,85)}=3.4$ p=3.80e-2 | $F_{(2,85)}=4.4$ p=1.53e-2 | $F_{(2,85)}=4.3$ p=1.67e-2 |
| ROI | $F_{(1,85)}=0.2$ p>0.05 | $F_{(1,85)}=63.7$ p=6.23e-12 | $F_{(1,85)}=4.4$ p=3.89e-2 | $F_{(1,85)}=132.5$ p<1.0e-16 |
| Group*ROI | $F_{(2,85)}=5.1$ | $F_{(2,85)}=8.5$ | $F_{(2,85)}=9.3$ | $F_{(2,85)}=3.7$ |

| | | | |
|---|---|---|---|
| | p=7.99e-3 | p=4.30e-4 | p=2.28e-4 | p=2.97e-2 |

### 3.4.2. Desynchronization ratio (DR)

At theta band, there were a significant main effect of Group and a significant effect of Group*ROI interaction, but no effect of ROI (Table 6). The post-hoc test in the frontal region showed that DR of AD subjects was significantly lower than those of CT subjects (p=7.76e-6, Fig. 3) and FTD subjects (p=7.24e-3, Fig. 3). The post-hoc test in the posterior region showed that DR of AD subjects was significantly lower than those of CT subjects (p=1.21e-4, Fig. 3) and FTD subjects (p=1.39e-2, Fig. 3).

At alpha band, there were significant main effects of Group and ROI as well as a significant effect of Group*ROI interaction (Table 6). The post-hoc test in the frontal region showed that DR of AD subjects was significantly lower than that of FTD subjects (p=2.49e-3, Fig. 3). The post-hoc test in the posterior region showed that DR of FTD subjects was significantly higher than those of AD subjects (p=3.30e-2, Fig. 3) and CT subjects (p=4.25e-3, Fig. 3).

At beta band, there were significant main effects of Group and ROI as well as a significant effect of Group*ROI interaction (Table 6). The post-hoc test in the frontal region showed that DR of FTD subjects was significantly higher than those of AD subjects (p=2.54e-5, Fig. 3) and CT subjects (p=4.19e-2, Fig. 3). The post-hoc test in the posterior region showed that DR of FTD subjects was significantly higher than those of AD subjects (p=1.69e-5, Fig. 3) and CT subjects (p=1.27e-4, Fig. 3).

At low gamma band, there were significant main effects of Group and ROI as well as a significant effect of Group*ROI interaction (Table 6). The post-hoc test in the frontal region showed that DR of AD subjects was significantly lower than those of FTD subjects (p=6.03e-6, Fig. 3) and CT subjects (p=3.26e-3, Fig. 3). The post-hoc test in the posterior region showed that DR of FTD subjects was significantly higher than those of AD subjects (p=3.47e-6, Fig. 3) and CT subjects (p=2.10e-2, Fig. 3). It also showed that DR of AD subjects was significantly lower than that of CT subjects (p=3.31e-2, Fig. 3).

**Table 6.** Desynchronization ratio (DR)

| | Theta band | Alpha band | Beta band | Low Gamma band |
|---|---|---|---|---|
| Group | $F_{(2,85)}=11.8$ p=2.97e-5 | $F_{(2,85)}=5.7$ p=4.92e-3 | $F_{(2,85)}=12.7$ p=1.53e-5 | $F_{(2,85)}=14.7$ p=3.38e-6 |
| ROI | $F_{(1,85)}=0.0$ p>0.05 | $F_{(1,85)}=68.0$ p=1.80e-12 | $F_{(1,85)}=11.0$ p=1.33e-3 | $F_{(1,85)}=178.0$ p<1.0e-16 |
| Group*ROI | $F_{(2,85)}=3.8$ p=2.72e-2 | $F_{(2,85)}=7.7$ p=8.40e-4 | $F_{(2,85)}=7.4$ p=1.09e-3 | $F_{(2,85)}=3.2$ p=4.50e-2 |

## 3.5. Functional network measures

### 3.5.1. Clustering coefficient

At theta band, there were a significant main effect of Group and a significant effect of Group*ROI interaction, but no effect of ROI (Table 7). The post-hoc test in the frontal region showed that CC of CT subjects was significantly higher than those of AD subjects (p=2.76e-4, Fig. 4) and FTD subjects (p=4.61e-3, Fig. 4). The post-hoc test in the posterior region showed no differences among groups (p>0.05).

At alpha band, there was a significant main effect of ROI, but no effect of Group, and no effect of Group*ROI interaction (Table 7). The post-hoc test in both frontal and posterior regions showed no differences among groups (p>0.05).

At beta band, there were a significant main effect of Group and a significant effect of Group*ROI interaction, but no effect of ROI (Table 7). The post-hoc test in the frontal region showed no differences among groups (p>0.05). The post-hoc test in the posterior region showed that CC of CT subjects was significantly lower than those of AD subjects (p=4.74e-2, Fig. 4) and FTD subjects (p=3.82e-5, Fig. 4). It also showed that CC of AD subjects was significantly lower than that of FTD subjects (p=2.97e-2, Fig. 4).

At low gamma band, there was a significant main effect of ROI, but no effect of Group, and no effect of Group*ROI interaction (Table 7). The post-hoc test in the frontal region showed no differences among groups (p>0.05). The post-hoc test in the posterior region showed that CC of FTD subjects was significantly higher than those of AD subjects (p=1.67e-2, Fig. 4) and CT subjects (p=1.56e-2, Fig. 4).

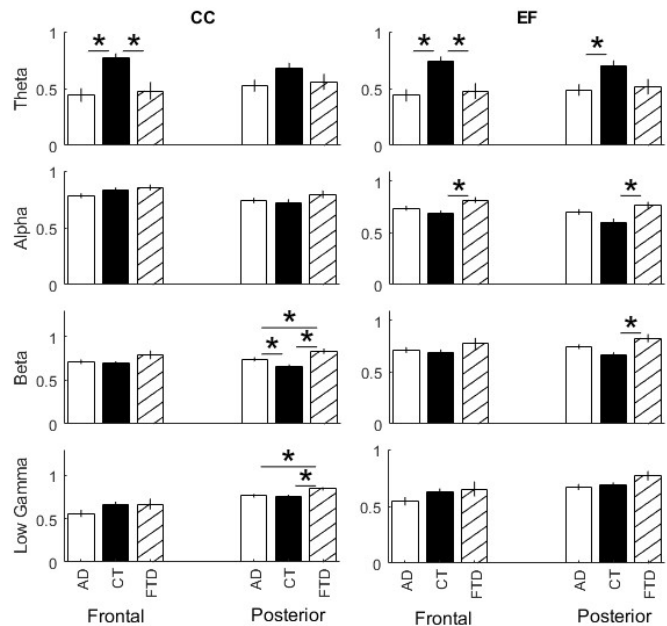

**Figure 4.** The clustering coefficient and network efficiency. Mean±SEM. Tukey's HSD post-hoc test. *: p<0.05.

**Table 7.** Clustering coefficient (CC)

|  | Theta band | Alpha band | Beta band | Low Gamma band |
|---|---|---|---|---|
| Group | F(2,85)=5.8 p=4.38e-3 | F(2,85)=1.9 p>0.05 | F(2,85)=5.8 p=4.18e-3 | F(2,85)=2.7 p>0.05 |
| ROI | F(1,85)=0.7 p>0.05 | F(1,85)=24.0 p=4.47e-6 | F(1,85)=0.9 p>0.05 | F(1,85)=50.5 p=3.45e-10 |
| Group*ROI | F(2,85)=5.6 p=4.99e-3 | F(2,85)=2.4 p>0.05 | F(2,85)=3.3 p=4.36e-2 | F(2,85)=2.5 p>0.05 |

*3.5.2. Network efficiency*

At theta band, there were a significant main effect of Group and a significant effect of Group*ROI interaction, but no effect of ROI (Table 8). The post-hoc test in the frontal region showed that EF of CT subjects was significantly higher than those of AD subjects (p=2.13e-4, Fig. 4) and FTD subjects (p=4.27e-3, Fig. 4). The post-hoc test in the posterior region showed that EF of CT subjects was significantly higher than that of AD subjects (p=1.04e-2, Fig. 4).

At alpha band, there were significant main effects of Group and ROI as well as a significant effect of Group*ROI interaction (Table 8). The post-hoc test in the frontal region showed that EF of FTD subjects was significantly higher than that of CT subjects (p=5.71e-3, Fig. 4). The post-hoc test in the posterior region showed that EF of FTD subjects was significantly higher than that of CT subjects (p=4.13e-3, Fig. 4).

At beta band, there were significant main effects of Group and ROI as well as a significant effect of Group*ROI interaction (Table 8). The post-hoc test in the frontal region showed no differences among groups (p>0.05). The post-hoc test in the posterior region showed that EF of FTD subjects was significantly higher than that of CT subjects (p=8.21e-4, Fig. 4).

At low gamma band, there were a significant main effect of ROI and a significant effect of Group*ROI interaction, but no effect of Group (Table 8). The post-hoc test in both frontal and posterior regions showed no differences among groups (p>0.05).

**Table 8.** Efficiency (EF)

|  | Theta band | Alpha band | Beta band | Low Gamma band |
|---|---|---|---|---|
| Group | F(2,85)=7.3 p=1.18e-3 | F(2,85)=5.4 p=6.18e-3 | F(2,85)=3.3 p=4.21e-2 | F(2,85)=2.0 p>0.05 |
| ROI | F(1,85)=1.9 p>0.05 | F(1,85)=94.4 p=1.93e-15 | F(1,85)=4.8 p=3.10e-2 | F(1,85)=74.0 p=3.42e-13 |
| Group*ROI | F(2,85)=7.1 p=1.38e-3 | F(2,85)=6.2 p=3.16e-3 | F(2,85)=6.5 p=2.31e-3 | F(2,85)=3.1 p=4.93e-2 |

### 3.6. Classification

*3.6.1. AD+FTD vs CT*

We compare AD+FTD with healthy control subjects to identify pathological vs. healthy state. As suggested by the results in the previous sections, the spectral measures, synchronized dynamics measures, and functional network measures may discriminate the disease states (AD and FTD) from the healthy state. In fact, we found that the following six measures (theta 1/f oscillatory power at T5, alpha 1/f oscillatory power at O1, alpha PSD at T5, theta synchronization index at F7-T6, and beta synchronization index at Pz-O2) can discriminate two states (AD+FTD vs CT) with a high accuracy. Table 9 demonstrated the accuracy, sensitivity, and specificity results by employing five machine learning classification algorithms. Table 10 presented the classifiers and the model hyper-parameters.

**Table 9.** Accuracy, sensitivity, and specificity results of five classification algorithms using 10-fold cross validation to discriminate AD+FTD with CT.

| AD+FTD vs CT | Accuracy (%) | Sensitivity (%) | Specificity (%) |
|---|---|---|---|
| Medium Tree | 79.6 | 84.6 | 69.3 |
| Linear Discriminat | 88.1 | 92.2 | 79.7 |
| Quadratic SVM | **92.4** | **95.3** | **86.6** |

| Fine KNN | 87.5 | 90.5 | 81.4 |
| Bagged Tree | 85.9 | 91.5 | 74.5 |

**Table 10.** Model Hyperparameters

| ML Algorithms | Medium Tree | Linear Discriminant | Quadratic SVM | Fine KNN | Bagged Tree |
|---|---|---|---|---|---|
| Hyperparameters | Max number of splits: 20

Split criterion: Gini's diversity index | Covariance structure: Full | Kernel function: Quadratic

Kernel scale: Auto

Box constraint level: 1

Multiclass coding: One-vs-One

Standardize data: Yes | Number of neighbors: 1

Distance metric: Euclidean

Distant weight: Equal

Standardize: Yes | Ensemble method: Bag

Leaner type: Decision tree

Maximum number of splits: 58

Number of learners: 30

Number of predictors to sample: Select All |

The best dementia diagnosis was achieved by Quadratic SVM with 92.4% accuracy, 95.3% sensitivity, and 86.6% specificity.

### 3.6.2. AD vs FTD

As suggested by the results in the previous sections, we observed that synchronized dynamics measures may discriminate between AD and FTD states. In fact, we found that the following four measures (beta DR at F7-P4, beta DR at F7-O2, beta DR Fz-T6, and low gamma DR at F7-O2) can discriminate two states (AD vs FTD) with a high accuracy. Table 11 demonstrated the accuracy, sensitivity, and specificity results by employing five machine learning classification algorithms.

**Table 11.** Accuracy, sensitivity, and specificity results of five classification algorithms using 10-fold cross validation to discriminate AD with FTD.

| AD vs FTD | Accuracy (%) | Sensitivity (%) | Specificity (%) |
|---|---|---|---|
| Medium Tree | 83.1 | 85.3 | 79.6 |
| Linear Discriminant | 74.4 | 80.8 | 64.4 |
| Quadractic SVM | 74.4 | 84.7 | 58.3 |
| Fine KNN | 81.4 | 89.2 | 69.1 |
| Bagged Tree | **89.7** | **96.1** | **79.6** |

The best discrimination between AD and FTD was achieved by the Bagged Tree with 89.7% accuracy, 96.1% sensitivity, and 79.6% specificity.

## 4. Discussion

The goal of the current study was to use dynamical system approach to explore the differences in the measures of resting state EEG between AD, FTD, and neurologically healthy controls to develop an integrative understanding of dynamics and organization of the neural functional connectivity in these conditions. We aimed not to validate specific measures, but rather to explore the neural dynamics using an array of dynamics measures that characterize related aspects of spatiotemporal dynamics of brain networks, and that can be taken together to infer the effect of dynamic organization on function, including behavior. Below, we summarize our observations, discuss the implications of functional activity on network organization and behavioral symptoms in AD and FTD, and conclude with a discussion of potential hypotheses suggested by our observations.

### 4.1 Summary of the results

Power spectral measures indicate a variety of changes in clinical population under study (see Sec 3.2), however, there appears to be one relatively consistent trend, that is exhibited by both estimates of the spectral power - 1/f oscillatory power and PSD. Alpha band power is higher in controls vs AD and FTD as indicated by both measures. As alpha frequency dominates EEG signal in awake humans, it is understood to play a crucial role in information integration across brain regions (Benwell et al., 2019; Palva & Palva, 2011). Thus, lowered alpha power in age-matched clinical cohorts compared to controls suggests likely deterioration of network dynamic state trajectories in AD and FTD. PSD indicates this for the posterior areas, while 1/f oscillatory power shows the same effect for both frontal and posterior regions in both groups. At the same time, spectral power tends to be lower in controls at theta band. While the details vary (see Sec 3.2), PSD indicates higher theta power in frontal areas for AD, while 1/f oscillatory power points to the higher theta power for AD in both posterior and frontal areas and for FTD in posterior area. As there is a variability of the outcomes across the bands, brain areas, and conditions, it may be not very easy to make very specific conclusions except that controls tend to have higher power in alpha band and lower power in theta band. This can potentially be an indicator of stronger local coherence in alpha band and weaker local coherence in theta band. These results also indicate that the ratio of theta to alpha power (also known as theta-alpha ratio (TAR)), for both AD and FTD would be higher than in controls. This is in line with earlier studies: an increased TAR, resulting from increased theta power and decreased alpha power, is observed in both AD and FTD when compared to controls, with differences in spatiotemporal patters between the patients with the two diagnoses (Wang et al., 2024). Moreover, in AD elevated TAR is associated with cognitive decline (van der Hiele et al., 2007). As FTD has several subtypes, those have been found to differ somewhat with regard to TAR indicators (e.g., Salcini et al., 2020).

Maximal Lyapunov exponent was used here as a measure of dynamic regularity of the neural activity. Lyapunov exponent characterizes the rate of divergence of the nearby trajectories in the phase space (Abarbanel, 1996; Kantz & Schreiber, 2003). That is, it informs on how quickly two very similar states in the network will grow apart in time. A larger Lyapunov exponent indicates faster divergence and, thus, less predictable, more irregular system dynamics. The Lyapunov exponent was observed to be statistically significantly higher in AD in posterior areas pointing to more irregular/chaotic network dynamics there.

We studied two measures of temporal dynamics. The first one is the synchronization (phase-locking) strength index; larger values of this index indicate stronger synchrony, pointing to the stronger temporal coordination of oscillatory activity. The second one is the desynchronization ratio. Higher values of the desynchronization ratio indicate the prevalence of short desynchronization events in partially synchronized oscillatory dynamics (more frequent transitions between desynchronized and synchronized states). The desynchronization ratio may vary independently of the phase-locking index (Ahn et al., 2011, 2014; Ahn & Rubchinsky, 2017; Nguyen & Rubchinsky, 2021, 2024), as the same synchrony strength may be achieved via high numbers of short desynchronizations or lower number of longer desynchronizations (or a variety of possibilities in between these two). AD showed lower synchrony strength than controls in theta band in both frontal and posterior areas. FTD showed higher synchrony strength in alpha and beta bands in posterior areas. FTD also showed higher low gamma band synchrony strength than that of AD in both frontal and posterior areas. In comparison with controls, AD has lower DR in theta and low gamma bands in both frontal and posterior areas, while FTD has higher DR in alpha, beta, and low gamma bands in posterior areas as well as higher DR in beta band in frontal area. Also, FTD has shown higher values of DR than AD at all frequency bands in both frontal and posterior areas. Thus, neural activity in AD appears to be less organized, less spatiotemporally coordinated, less interdependent (at least at the lower frequency bands) than controls. On the other hand, neural activity in FTD appears to be more organized (except in the theta band – the 4-8 Hz portion of the spectrum) than controls. Furthermore, neural activity in FTD is characterized by more frequent dynamic transitions between desynchronization and synchronization. The neural dynamics demonstrate that FTD is more synchronized and organized than AD.

We also considered two network-theoretic measures of organization of the functional brain dynamics, clustering and efficiency. Of course, every functional network has some anatomical networks underlying it, but the properties of the organization of the functional networks first inform us about neural activity rather than underlying anatomical structure. The clustering coefficient (CC) measures the tendency to form locally dense connectivity. On the other hand, efficiency (which is reciprocal to the shortest path length) characterizes the efficiency of the connectivity (and thus, presumably, information transfer) in the network overall. We observed that functional networks in both AD and FTD exhibited lower clustering coefficient and network efficiency in comparison with controls in the low frequency (theta frequency band), primarily in frontal areas. At the higher frequencies (beta and lower gamma, and also alpha for network efficiency), functional networks in FTD exhibited higher functional connectivity and efficiency compared to those of controls.

Finally, we used machine learning classification to complement conventional statistical analyses in assessing the salience of proposed measures for characterizing neural dynamics in the two clinical cohorts. Unlike parametric statistical methods, machine learning approaches are not bound by specific distributional assumptions. Our findings indicate that six salient dynamics measures - theta 1/f oscillatory power at T5, alpha 1/f oscillatory power at O1, alpha PSD at T5, theta synchronization index at F7-T6, and beta synchronization index at Pz-O2 - allowed to distinguish between clinical and control groups with 92.4% accuracy using Quadratic SVM algorithm. Additionally, another set of four measures - beta DR at F7-P4, beta DR at F7-O2, beta DR at Fz-T6, and low gamma DR at F7-O2 - differentiated AD from FTD with 89.7% accuracy using a Bagged Tree classifier. While our results showed slightly lower accuracy for AD and FTD vs. CT classification as compared to solely diagnostics-focused work (Rostamikia et al., 2024), the classification performance for AD vs. FTD based on the measures of dynamics was, in fact, slightly improved.

## 4.2 From neural activity measures to the understanding of the functional netowrks dynamics in AD and FTD

There are multiple differences between AD and FTD (and control subjects), and these differences span different brain areas, spectral bands, and neural activiy measures. The summary of the results presented above suggests several trends described by convergent findings in spatiotemporal measures regarding neural dynamics of AD and FTD.

Spectral power measures of EEG point to more coherent, more coordinated local activity in AD and in FTD in both theta and higher (lower gamma) bands with respect to controls. A lowering of alpha power in clinical cohorts, which typically facilitates cross-regional communication, may lead to a shift toward more locally structured neural activity. Particularly, this would be reflected in increased coherence and coordination of local activity particularly in the low gamma band, in FTD patients as compared to controls. However, lowering of alpha power in AD and FTD patients likely reflects weakening of the brain's ability to integrate information across regions; this shift could underlie the cognitive deficits reflected in MMSE test performance in AD and FTD groups, which underscores the impact of disrupted neural dynamics on cognitive function.

Synchronous dynamics measures suggest that neural acrtivity in AD tends to be less synchronized with longer desycnrhonizations, while FTD activity tends to be more synchronized with shorter desynchronization. Regularity measure points to a more chaotic and unstable dynamics in AD. And network measures also point to, generally speaking, less locally connected and less globally connected funcitonal networks in AD, while functional networks in FTD are less locally and globally connected in fontal areas in slow rhythms and are more locally connected and more globally connected otherwise. This picture suggests that AD neural activity may be less coordinated and less connected across areas, and more random, while FTD shows more coordinated neural activity (except slow frontal activity). In other words, despite some similarity in neurological deficits, AD and FTD may, in some ways, be on two different sides of healthy brain: AD is less spatiotemporally coordinated and more random, FTD is more connected and more coordinated. Deviations from normal in both directions may lead to neurological deficits, which are specific to each of the disorders.

These findings align with the observed differences in MMSE performance between AD and FTD in our patient groups, as well as MMSE and Activities of Daily Living (ADL) tests in the literature. Frontotemporal dementia (FTD) patients tend to have MMSE scores than Alzheimer's disease (AD) patients, even when they exhibit similar levels of impairment in ADL (Lin et al., 2021). This difference arises as AD primarily affects memory, which is a key domain assessed by the MMSE. The less coordinated, more random neural activity in AD, reflected in longer desynchronizations and unstable dynamics, corresponds to its profound impact on memory and cognition, as captured by low MMSE scores in this group. In contrast, FTD is characterized by more subtle impairments in emotional regulation, social behavior, and executive function, which are not assessed by the MMSE. While these deficits impact decision-making and social interactions, they do not lead to functional decline in basic daily activities. As a result, FTD patients may perform relatively well on the MMSE, despite noticeable behavioral and executive dysfunction. The more synchronized, structured neural activity observed in FTD may explain why executive and behavioral impairments in FTD do not immediately translate into functional deficits in daily living, as the large-scale structure of neural dynamics supporting, for example, motor functions can be preserved in FTD, even as fine spatiotemporal dependencies are disrupted.

We would like to note that the situation with relatively minor and not very consistent differences between conditions is, perhaps, not surprising if the neural dynamics is not very different (Ahn et al., 2018; Malaia et al., 2020). In that respect, the more robust differences in the desynchronization ratio rather than the syncrhonization index are expected, given how DR tends to be more sensitive to small changes in the dynamics, and now changes in DR may occur earlier than changes in the overall synchrony (as seen in other experimental systems, cf. Ahn et al., 2014, 2018).

Finally, a recent study by Liao et al. (2025) used machine learning classification techniques to identify differences in task-based ERP data between AD patients and controls. When the brain is engaged in a working memory task, EEG-based ERPs may be more likely to reveal distinctions between AD and control groups, potentially aiding in the identification of prodromal AD. However, the task-based alterations in quantitative EEG detected by machine learning ultimately stem from underlying abnormalities in the state trajectory traversal of brain dynamics caused by neurodegeneration. In contrast, our study focuses on resting-state EEG, allowing us to characterize these abnormalities more fundamentally in terms of the neural dynamics of EEG-based networks when the brain is not engaged in an external task. This approach provides a clearer picture of the intrinsic disruptions in neural coordination and connectivity associated with AD.

## 4.3 Conclusion

The analysis of neural dynamics in functional brain networks in AD and FTD presented here emphasized the complexity of these disorders, which cannot be understood through a single measure of neural activity. Different spectral bands and brain regions may show distinct differences from control group, reflecting the specific roles neural frequencies and brain areas play in different types of information processing. However, by integrating multiple dynamical measures, we can identify common underlying features that represent the specific neurophysiological changes that characterize AD and FTD.

Our findings suggest that AD and FTD may represent opposite changes from normal brain function in terms of the spatiotemporal coordination of neural activity, with AD exhibiting less coordinated, less synchronized, and more random activity, while FTD shows excessive coordination and synchronization (except in slow frontal rhythms). This broad conclusion is not derived from any single measure; rather, it emerges from an approach that considers multiple interacting dynamical features of neural activity. While the applicability of our analysis to prodromal clinical diagnostics may be limited, since we consider advanced stages of both AD and FTD, modeling nonlinear dynamics of neural activity contributes to a deeper understanding of the fundamental neurophysiological principles underlying the symptoms of these distinct yet related disorders.

**Acknowledgements**
Supported by AMS-Simons Grant for PUI (SA) and NSF 1734938 (EAM).

**Author contributions**
SA, EM, LR: conceptualized and designed the study: SA performed numerical analysis, SA, EM, LR analyzed the results: SA, EM, LR wrote the manuscript.

## Declaration of competing interest

The authors declare that they have no known competing financial interests or personal relationships that could have appeared to influence the work reported in this paper.

## References

Abarbanel, H. (1996). *Analysis of observed chaotic data*. Springer Science & Business Media.

Agosta, F., Spinelli, E. G., Basaia, S., Cividini, C., Falbo, F., Pavone, C., Riva, N., Canu, E., Castelnovo, V., Magnani, G., Caso, F., Caroppo, P., Prioni, S., Villa, C., Tremolizzo, L., Appollonio, I., Silani, V., Josephs, K. A., Whitwell, J., & Filippi, M. (2023). Functional Connectivity From Disease Epicenters in Frontotemporal Dementia. *Neurology*, *100*(22). https://doi.org/10.1212/WNL.0000000000207277

Ahn, S., Park, C., & Rubchinsky, L. L. (2011). Detecting the temporal structure of intermittent phase locking. *Physical Review E*, *84*(1), 016201. https://doi.org/10.1103/PhysRevE.84.016201

Ahn, S., & Rubchinsky, L. L. (2013). Short desynchronization episodes prevail in synchronous dynamics of human brain rhythms. *Chaos: An Interdisciplinary Journal of Nonlinear Science*, *23*(1), 013138. https://doi.org/10.1063/1.4794793

Ahn, S., & Rubchinsky, L. L. (2017). Potential mechanisms and functions of intermittent neural synchronization. *Frontiers in Computational Neuroscience*, *11*, 44. https://doi.org/10.3389/fncom.2017.00044

Ahn, S., Rubchinsky, L. L., & Lapish, C. C. (2014). Dynamical Reorganization of Synchronous Activity Patterns in Prefrontal Cortex–Hippocampus Networks During Behavioral Sensitization. *Cerebral Cortex*, *24*(10), 2553–2561. https://doi.org/10.1093/cercor/bht110

Ahn, S., Zauber, S. E., Worth, R. M., Witt, T., & Rubchinsky, L. L. (2018). Neural synchronization: Average strength vs. temporal patterning. *Clinical Neurophysiology*, *129*(4), 842–844. http://doi.org/10.1016/j.clinph.2018.01.063

Alzheimer's Association. (2024). 2024 Alzheimer's disease facts and figures. *Alzheimer's & Dementia, 20*(5), 3708-3821. https://doi.org/10.1002/alz.13809

Arbabyazd, L., Petkoski, S., Breakspear, M., Solodkin, A., Battaglia, D., & Jirsa, V. (2023). State-switching and high-order spatiotemporal organization of dynamic functional connectivity are disrupted by Alzheimer's disease. *Network Neuroscience (Cambridge, Mass.)*, *7*(4), 1420–1451. https://doi.org/10.1162/netn_a_00332

Benwell, C. S. Y., London, R. E., Tagliabue, C. F., Veniero, D., Gross, J., Keitel, C., & Thut, G. (2019). Frequency and power of human alpha oscillations drift systematically with time-on-task. *Neuroimage*, *192*, 101–114. https://doi.org/10.1016/j.neuroimage.2019.02.067

Bloom, G. S. (2014). Amyloid-β and tau: The trigger and bullet in Alzheimer disease pathogenesis. *JAMA Neurology*, *71*(4), 505–508. https://doi.org/10.1001/jamaneurol.2013.5847

Bullmore, E., & Sporns, O. (2012). The economy of brain network organization. *Nature Reviews Neuroscience*, *13*(5), 336–349. https://doi.org/10.1038/nrn3214

Chu, M., Jiang, D., Li, D., Yan, S., Liu, L., Nan, H., Wang, Y., Wang, Y., Yue, A., & Ren, L. (2024). Atrophy network mapping of clinical subtypes and main symptoms in frontotemporal dementia. *Brain*, awae067. https://doi.org/10.1093/brain/awae067

Ferreira, L. K., Lindberg, O., Santillo, A. F., & Wahlund, L. (2022). Functional connectivity in behavioral variant frontotemporal dementia. *Brain and Behavior*, *12*(12), e2790. https://doi.org/10.1002/brb3.2790

Guzmán-Vélez, E., Diez, I., Schoemaker, D., Pardilla-Delgado, E., Vila-Castelar, C., Fox-Fuller, J. T., Baena, A., Sperling, R. A., Johnson, K. A., Lopera, F., Sepulcre, J., & Quiroz, Y. T. (2022). Amyloid-β and tau pathologies relate to distinctive brain dysconnectomics in preclinical autosomal-dominant Alzheimer's disease. *Proceedings of the National Academy of Sciences*, *119*(15), e2113641119. https://doi.org/10.1073/pnas.2113641119

Hurtado, J. M., Rubchinsky, L. L., & Sigvardt, K. A. (2004). Statistical method for detection of phase-locking episodes in neural oscillations. *Journal of Neurophysiology*, *91*(4), 1883–1898. https://doi.org/10.1152/jn.00853.2003

Kantz, H., & Schreiber, T. (2003). *Nonlinear time series analysis*. Cambridge university press. 2nd ed.

Latora, V., & Marchiori, M. (2001). Efficient Behavior of Small-World Networks. *Physical Review Letters*, *87*(19), 198701. https://doi.org/10.1103/PhysRevLett.87.198701

Liao, K., Martin, L. E., Fakorede, S., Brooks, W. M., Burns, J. M., & Devos, H. (2025). Machine learning based on event-related oscillations of working memory differentiates between preclinical Alzheimer's disease and normal aging. *Clinical Neurophysiology*, *170*, 1–13. https://doi.org/10.1016/j.clinph.2014.11.013

Lin, N., Gao, J., Mao, C., Sun, H., Lu, Q., & Cui, L. (2021). Differences in Multimodal EEG and Clinical Correlations Between Alzheimer's Disease and Frontotemporal Dementia. *Frontiers in Neuroscience*, 15,687053. https://doi.org/10.3389/fnins.2021.607053

Malaia, E. A., Ahn, S., & Rubchinsky, L. L. (2020). Dysregulation of temporal dynamics of synchronous neural activity in adolescents on autism spectrum. *Autism Research*, *13*(1), 24–31. https://doi.org/10.1002/aur.2219

Mann, D. M. A., & Snowden, J. S. (2017). Frontotemporal lobar degeneration: Pathogenesis, pathology and pathways to phenotype. *Brain Pathology*, *27*(6), 723–736. https://doi.org/10.1111/bpa.12486

McDonnell, J., Murray, N. P., Ahn, S., Clemens, S., Everhart, E., & Mizelle, J. C. (2021). Examination and Comparison of Theta Band Connectivity in Left-and Right-Hand Dominant Individuals throughout a Motor Skill Acquisition. *Symmetry*, *13*(4), 728. https://doi.org/10.3390/sym13040728

Micheloyannis, S., Pachou, E., Stam, C. J., Vourkas, M., Erimaki, S., Tsirka, V. (2006). Using graph theoretical analysis of multi channel EEG to evaluate the neural efficiency hypothesis. *Neuroscience Letters*, 402(3), 273-277. https://doi.org/10.1016/j.neulet.2006.04.006

Miltiadous, A., Tzimourta, K. D., Afrantou, T., Ioannidis, P., Grigoriadis, N., Tsalikakis, D. G., Angelidis, P., Tsipouras, M. G., Glavas, E., Giannakeas, N., & Tzallas, A. T. (2023). A Dataset of Scalp EEG Recordings of Alzheimer's Disease, Frontotemporal Dementia and Healthy Subjects from Routine EEG. *Data*, *8*(6), Article 6. https://doi.org/10.3390/data8060095

Nguyen, Q.-A., & Rubchinsky, L. L. (2021). Temporal patterns of synchrony in a pyramidal-interneuron gamma (PING) network. *Chaos (Woodbury, N.Y.)*, *31*(4), 043134. https://doi.org/10.1063/5.0042451

Nguyen, Q.-A., & Rubchinsky, L. L. (2024). Synaptic effects on the intermittent synchronization of gamma rhythms. *Cognitive Neurodynamics*, 18, 3821-3837. https://doi.org/10.1007/s11571-024-10150-9

Palva, S., & Palva, J. M. (2011). Functional Roles of Alpha-Band Phase Synchronization in Local and Large-Scale Cortical Networks. *Frontiers in Psychology*, 2. https://doi.org/10.3389/fpsyg.2011.00204

Park, C., Worth, R. M., & Rubchinsky, L. L. (2010). Fine Temporal Structure of Beta Oscillations Synchronization in Subthalamic Nucleus in Parkinson's Disease. *Journal of Neurophysiology*, 103(5), 2707–2716. https://doi.org/10.1152/jn.00724.2009

Pikovsky, A., Rosenblum, M., & Kurths, J. (2001). *Synchronization: A universal concept in nonlinear sciences*. Cambridge University Press.

Rostamikia, M., Sarbaz, Y., & Makouei, S. (2024). EEG-based classification of Alzheimer's disease and frontotemporal dementia: A comprehensive analysis of discriminative features. *Cognitive Neurodynamics*, 18(6), 3447–3462. https://doi.org/10.1007/s11571-024-10152-7

Salcini, C., Cebi, M., Sari, A. B., Tanridag, O., & Tarhan, N. (2020). Quantitative EEG differences in subtypes of frontotemporal dementia. *Psychiatry and Clinical Psychopharmacology*, 30(2), 182–185. https://doi.org/10.5455/pcp.20200215073554

Stam, C., & Reijneveld, J. C. (2007). Graph theoretical analysis of complex networks in the brain. *Nonlinear Biomedical Physics*, 1(1), 3. https://doi.org/10.1186/1753-4631-1-3

van der Hiele, K., Vein, A. A., Reijntjes, R. H. a. M., Westendorp, R. G. J., Bollen, E. L. E. M., van Buchem, M. A., van Dijk, J. G., & Middelkoop, H. a. M. (2007). EEG correlates in the spectrum of cognitive decline. *Clinical Neurophysiology*, 118(9), 1931–1939. https://doi.org/10.1016/j.clinph.2007.05.070

Wang, Z., Liu, A., Yu, J., Wang, P., Bi, Y., Xue, S., Zhang, J., Guo, H., & Zhang, W. (2024). The effect of aperiodic components in distinguishing Alzheimer's disease from frontotemporal dementia. *GeroScience*, 46(1), 751–768. https://doi.org/10.1007/s11357-023-01041-8

Watts, D. J., & Strogatz, S. H. (1998). Collective dynamics of 'small-world' networks. *Nature*, 393(6684), 440–442. https://doi.org/10.1038/30918

Wen, H., & Liu, Z. (2016). Separating fractal and oscillatory components in the power spectrum of neurophysiological signal. *Brain Topography*, 29, 13–26. https://doi.org/10.1007/s10548-015-0448-0

Zheng, X., Wang, B., Liu, H., Wu, W., Sun, J., Fang, W., Jiang, R., Hu, Y., Jin, C., & Wei, X. (2023). Diagnosis of Alzheimer's disease via resting-state EEG: Integration of spectrum, complexity, and synchronization signal features. *Frontiers in Aging Neuroscience*, 15, 1288295. https://doi.org/10.3389/fnagi.2023.1288295